\documentclass[11pt]{article}
\usepackage{moriond,epsfig}

\def\Journal#1#2#3#4{{#1} {\bf #2}, #3 (#4)}

\def\PRL{\em Phys. Rev. Lett.}
\def\PRD{{\em Phys. Rev.} D}

\def\be{\begin{equation}}
\def\ee{\end{equation}}
\def\bea{\begin{eqnarray}}
\def\eea{\end{eqnarray}}

\begin{document}
\vspace*{4cm}
\title{Rate-Only Analysis with Reactor-Off Data in Double Chooz}

\author{ P. Novella }

\address{ APC-CNRS,\\
10, rue Alice Domon et Léonie Duquet,
75205 Paris, France}

\maketitle\abstracts{
Among ongoing reactor-based experiments, Double Chooz is unique in obtaining data when all reactor cores are brought down for maintenance. These reactor-off data allow for a clean measurement of the backgrounds of the experiment, thus being of uppermost importance for the $\theta_{13}$ oscillation analysis. While the oscillation results published by the collaboration in 2011 and 2012 rely on background models derived from reactor-on data, in this analysis we present an independent study based on the handle provided by 7.53 days of reactor-off data. A global fit to both $\theta_{13}$ and the total background is performed by analyzing the observed neutrino rate as a function of the non-oscillated expected rate for different reactor power conditions. The result presented in this study is fully consistent with the one already published by Double Chooz. As they both yield almost the same precision, this work stands as a prove of the reliability of the background estimates and the oscillation analysis of the experiment.}

\section{Reactor Rate Modulation Analysis}

In order to measure the mixing angle $\theta_{13}$ by means of reactor neutrino experiments, the observed $\bar{\nu}_e$ candidates rate ($R^{obs}$) can be confronted with the expected one ($R^{exp}$) in absence of oscillation. As Double Chooz (DC) data have been taken for different reactor power (P$_{th}$) conditions, the rate comparison can be done for different expected averaged rates. In particular, there are three well defined configurations as far as the operation of the two involved reactors is concerned: 1) the two reactors are on (2-On data), 2) one of the reactors is off (1-Off reactor data), and 3) both reactors are off (2-Off reactor data). For the 1-Off and 2-Off reactor data, the expected neutrino rate takes into account the residual neutrinos ($ R^{r-\nu} $) generated after the cores are brought down. In the presented study, the data sample in \cite{dc2012} is used, along with an extra 2-Off sample collected in 2012 \cite{offoff}, which increases the total 2-Off data taking time up to 7.53 days.

\begin{figure}[htbp]
\begin{center}
\includegraphics[width=60mm]{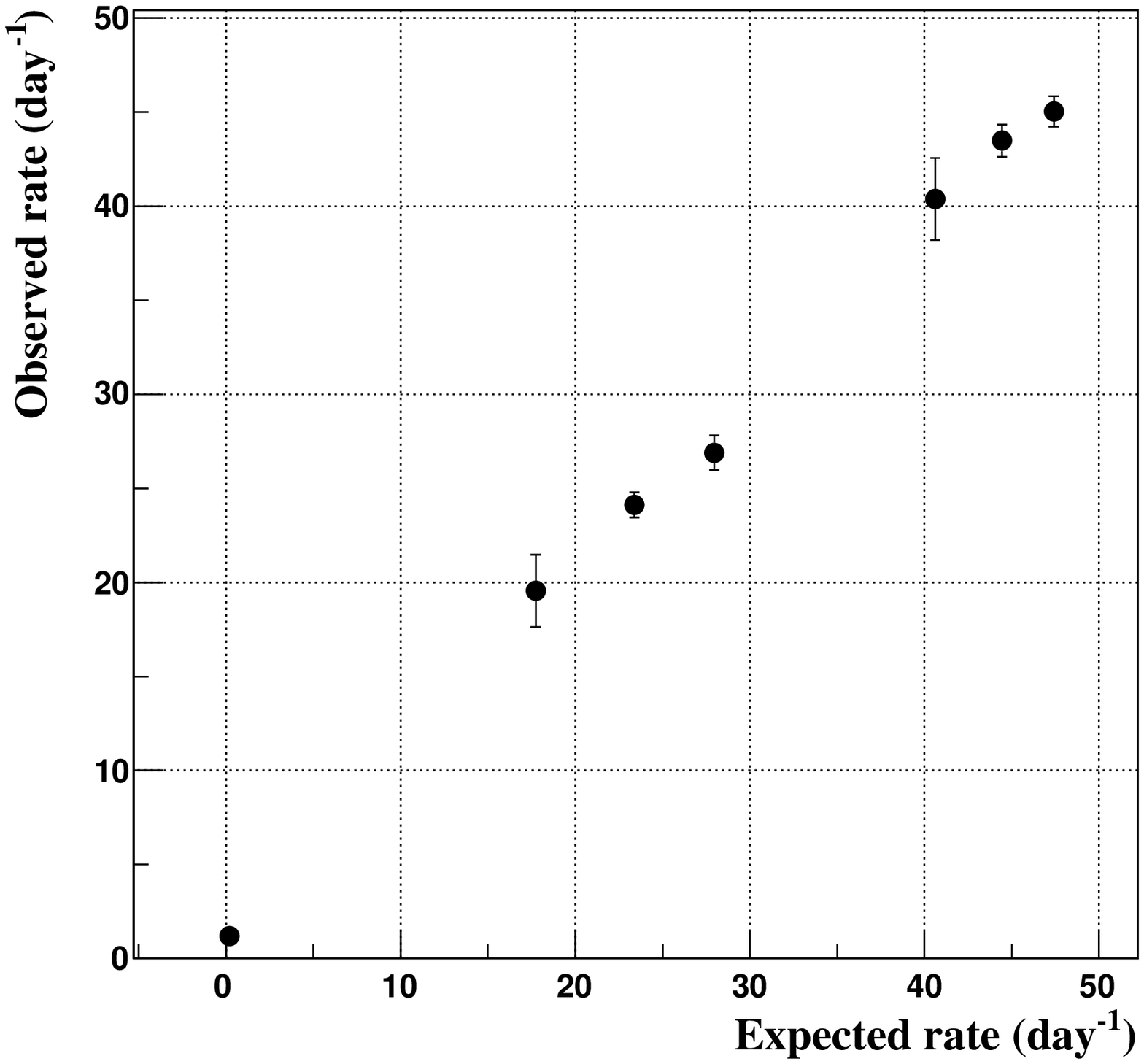}
\includegraphics[width=60mm]{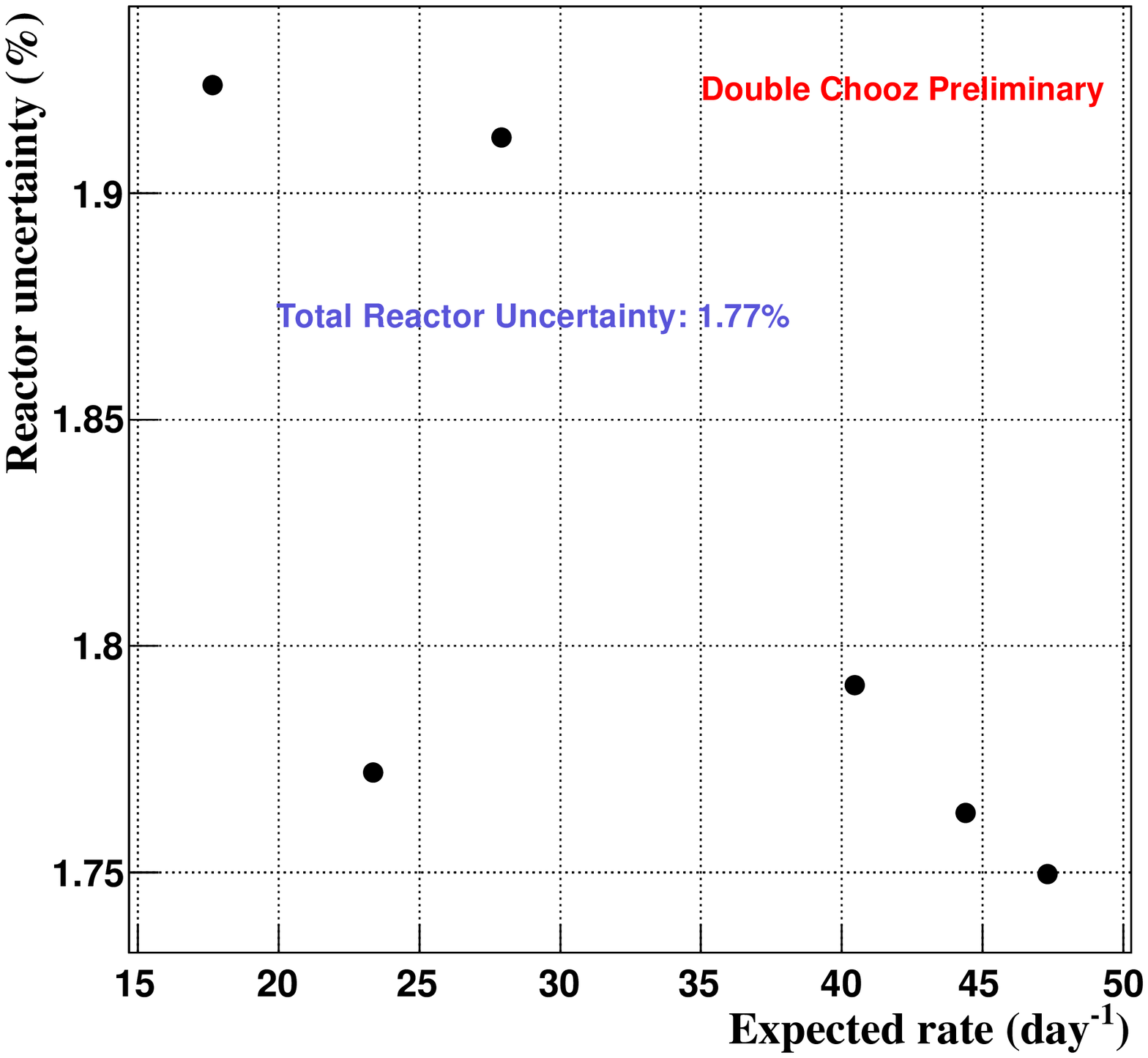}
\end{center}
\caption{\label{fig:obsvsexp}  Left: Observed versus expected neutrino candidate rate. Right: reactor-related uncertainties on the expected rate.}
\end{figure}

From the comparison between $R^{exp}$ and $R^{obs}$, as shown in left panel of Fig.~\ref{fig:obsvsexp}, both the value of $\theta_{13}$ and the total background rate $B$ can be derived. The correlation of the expected and observed rates follows a linear model parametrized by $\theta_{13}$ and $B$:

\begin{equation}
\label{eq:model} 
f(R^{exp})=B+\left(1-\sin^2(2\theta_{13})\cdot\alpha_{osc}\right)\cdot R^{exp}
\end{equation}

\noindent where $\alpha_{osc}$ is the average disappearance coefficient, $<\sin^2(\Delta m^2 L/4E)>$, computed by means of simulations to be 0.509. Fitting the data points in left plot of Fig.~\ref{fig:obsvsexp} to the above model provides a direct measurement of the mixing angle and the total background rate, along with their corresponding uncertainties. As the fitted values obtained with this technique are not based on any a priori assumption on the background, the result is independent to that one obtained in \cite{dc2012,dc2011}, where a background model, extracted from reactor-on data, is fitted along with the mixing angle. The accuracy and precision on the fitted value of $B$, as well as the accuracy on $\theta_{13}$, relies mostly on the 2-Off reactor data, as this sample provides a powerful lever arm for the fit.

\section{Systematic uncertainties}

There are three sources of systematics to be accounted for in the rate-only analysis: 1) detection efficiency ($\sigma_d$), 2) residual $\bar{\nu}_e$ prediction in reactor-off data ($\sigma_{\nu}$), and 3) $\bar{\nu}_e$ prediction in reactor-on data ($\sigma_r$). The detection efficiency systematics are listed in \cite{dc2012}, from where the total uncertainty $\sigma_d$ is derived to be 1.005\%. The rate of residual neutrinos and the associated uncertainty has been computed as described in \cite{offoff}, and a $\sigma_{\nu}$=30\% uncertainty is assigned to $R^{r-\nu}$ in all reactor-off periods. Finally, a dedicated study has been performed in order to estimate $\sigma_r$ as a function of the thermal power.

To a good approximation, all sources of reactor-related systematics are independent of P$_{th}$, with the exception of the uncertainty on P$_{th}$ itself, $\sigma_{P_{th}}$. This error is 0.5\% \cite{dc2012} when the reactors are running at full power, but it increases as P$_{th}$ decreases.  In  \cite{dc2012}, $\sigma_{P_{th}}$ is assumed to be 0.5\% for all data. This is a very good approximation when one integrates over all the data taking sample, and consequently all reactor operation conditions, as more than 90\% of the data are taken at full reactor power. However, in the current analysis this is not a valid approximation as it relies on data taken at different reactor powers. In order to compute $\sigma_{P_{th}}$ for different P$_{th}$ below the nominal one, a model assuming some constant systematic shift in the power plus a small contribution linear in the power is used. The model is fitted to a sample of measurements provided by EdF (company operating the Chooz nuclear plant), and the outcome is used to compute the errors in $R^{exp}$, as shown in the right panel of Fig.~\ref{fig:obsvsexp}. The total error $\sigma_r^i$ (where $i$ stands for each data point) ranges from 1.75\% (reactors operating at full power) to 1.92\% (one or two reactors not at full power).


\section{Oscillation and background results}

The $R^{obs}$ fit is based on a standard $\chi^2$ minimization. The $\chi^2$ definition is divided into three different terms: $\chi^2 = \chi^2_{on} + \chi^2_{off} + \chi^2_{pull}$, where $\chi^2_{on}$ stands for reactor-on and 1-Off reactor data, $\chi^2_{off}$ stands for 2-Off reactor data, and $\chi^2_{pull}$  accounts for the systematic uncertainties. Assuming gauss-distributed errors for the data points involving at least one reactor on, $\chi^2_{on}$ is built as follows:

\begin{equation}
\label{eq:chi2on}
\chi^2_{on} = \sum_i^{N} \frac{ \left(R_{i}^{obs} - f(R_{i}^{exp})\cdot[1+\alpha^d + \alpha_i^{r} + w_i\cdot\alpha_i^{\nu}]\right)^2}{ \sigma_{stat}^2} 
\end{equation}

\noindent where $N$ stands for the number of averaged rates (6), and where $\alpha^d$, $\alpha^r_i$ and $\alpha^{\nu}$ stand for pulls associated to the detection, reactor-on and reactor-off systematics, respectively. The fraction of residual neutrinos $w_i$ in each data point is defined as $
w_{i} = R_{i}^{r-\nu}/(R_{i}^{exp}+B)$.

Due to the low statistics in the 2-off reactor period, the corresponding error in $R^{obs}$ is considered as poisson-distributed like. As a consequence, $\chi^2_{off}$ is defined as a binned Poisson likelihood which follows a $\chi^2$ distribution:

\begin{equation}
\label{eq:chi2off}
\chi^2_{off} = 2 \left( R^{obs}\cdot T_{off}\cdot \textrm{ln} \frac{R^{obs}\cdot T_{off}}{K\cdot[1+\alpha^d+w_{i}\cdot\alpha^{\nu}]}+K\cdot[1+\alpha^d+w_{i}\cdot\alpha^{\nu}] - R^{obs}\cdot T_{off}\right)
\end{equation}

\noindent where $T_{off}$ is the live time of the 2-Off data sample, and $K$ is the number of expected events ($B+R^{r-\nu}\cdot T_{off}$). Finally, the $\chi^2_{pull}$ incorporates the penalty terms corresponding to $\sigma_r$, $\sigma_{d}$ and $\sigma_{\nu}$:

\begin{equation}
\chi^2_{pull} = (\frac{\alpha^d}{\sigma_d})^2 + \sum^N_i( \frac{\alpha_i^r}{\sigma_i^r} )^2 +  (\frac{\alpha^{\nu}}{\sigma_{\nu}})^2
\end{equation}

\begin{figure}[htbp]
\begin{center}
\includegraphics[width=60mm]{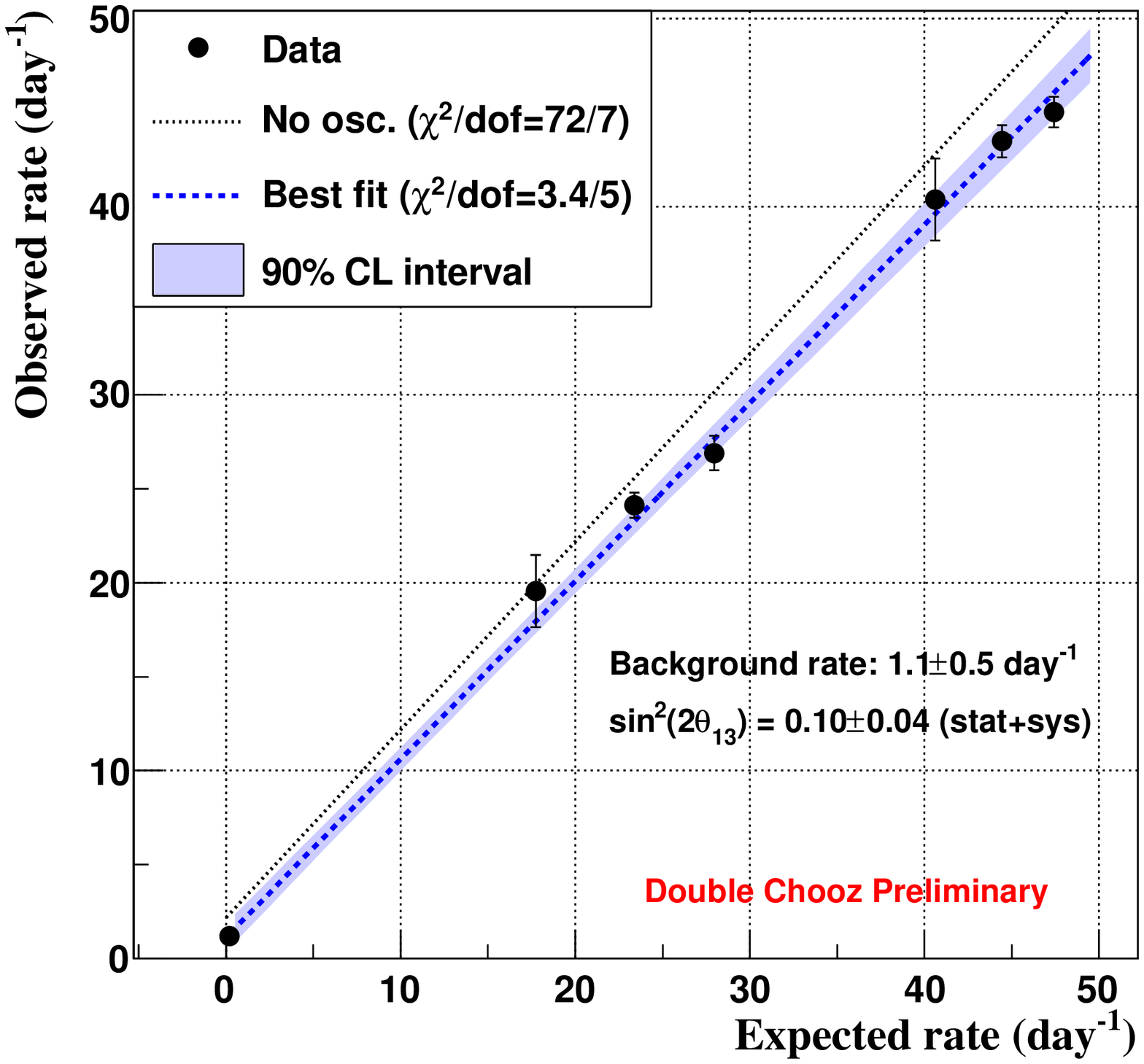}
\includegraphics[width=60mm]{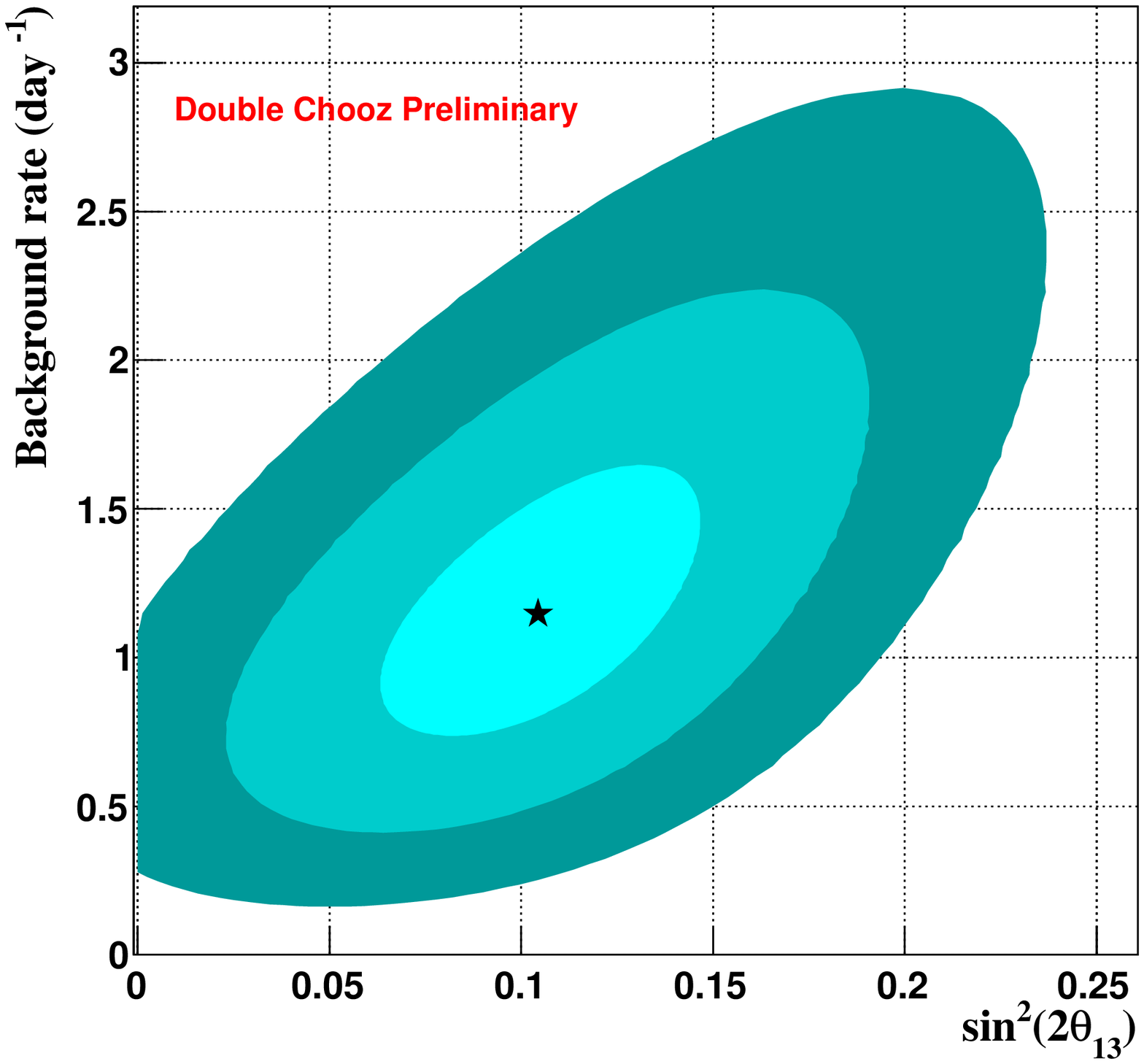}
\end{center}
\caption{\label{fig:fit1} Observed versus expected candidates rate fit. Left: (sin$^2(2\theta_{13})$,$B$) fit, superimposed to the null oscillation hypothesis. Right: 1, 2 and 3$\sigma$ ($\theta_{13}$,$B$) contour plot.}
\end{figure}

The outcome of the fit can bee seen in Fig.~\ref{fig:fit1}. The best fit values for sin$^{2}(2\theta_{13})$ and the total background rate are 0.10$\pm$0.04 and 1.1$\pm$0.5 events/day, respectively, with a $\chi^2$/dof of 3.4/5. Left plot of Fig.~\ref{fig:fit1} shows the fit and the corresponding 90\% confidence interval, superimposed to the null oscillation hypothesis assuming the background estimates in \cite{dc2012}. The 1, 2 and 3$\sigma$ (sin$^2(2\theta_{13})$,$B$) contour plot is also presented in right plot of the same Fig. As expected from the poisson treatement of the 2-Off data, the error on the background is constrained to positive values. These results are fully compatible with the ones obtained in \cite{dc2012}, yielding almost the same precision. 

While the best fit value of the total background rate depends on the neutrino candidates selection cuts, the one of $\theta_{13}$ must be independent. The current analysis can be performed for two different set of cuts: the ones applied in the first DC oscillation analysis \cite{dc2011}, and the ones applied in the second analysis \cite{dc2012}. The difference among these two selections relies on the use of the outer-veto and the showering muon veto, which are not applied in \cite{dc2011}, thus increasing the contamination of cosmogenic and correlated background events in the neutrino candidates sample. For the latter selection, the fit yields sin$^2(2\theta_{13})$=0.12$\pm$0.05 and $B$=2.9$\pm$0.6 events/day. As expected, the total background rate is larger that the one presented above, while the central value of $\theta_{13}$ is fully consistent. The precision on $B$ (relative error) is improved consistently with the fact that the statistical power of the 2-Off data is enhanced: while the selection in \cite{dc2012} yields 8 events in the 2-Off data sample, 21 events are selected with the one in \cite{dc2011}. 

\begin{figure}[htbp]
\begin{center}
\includegraphics[width=60mm]{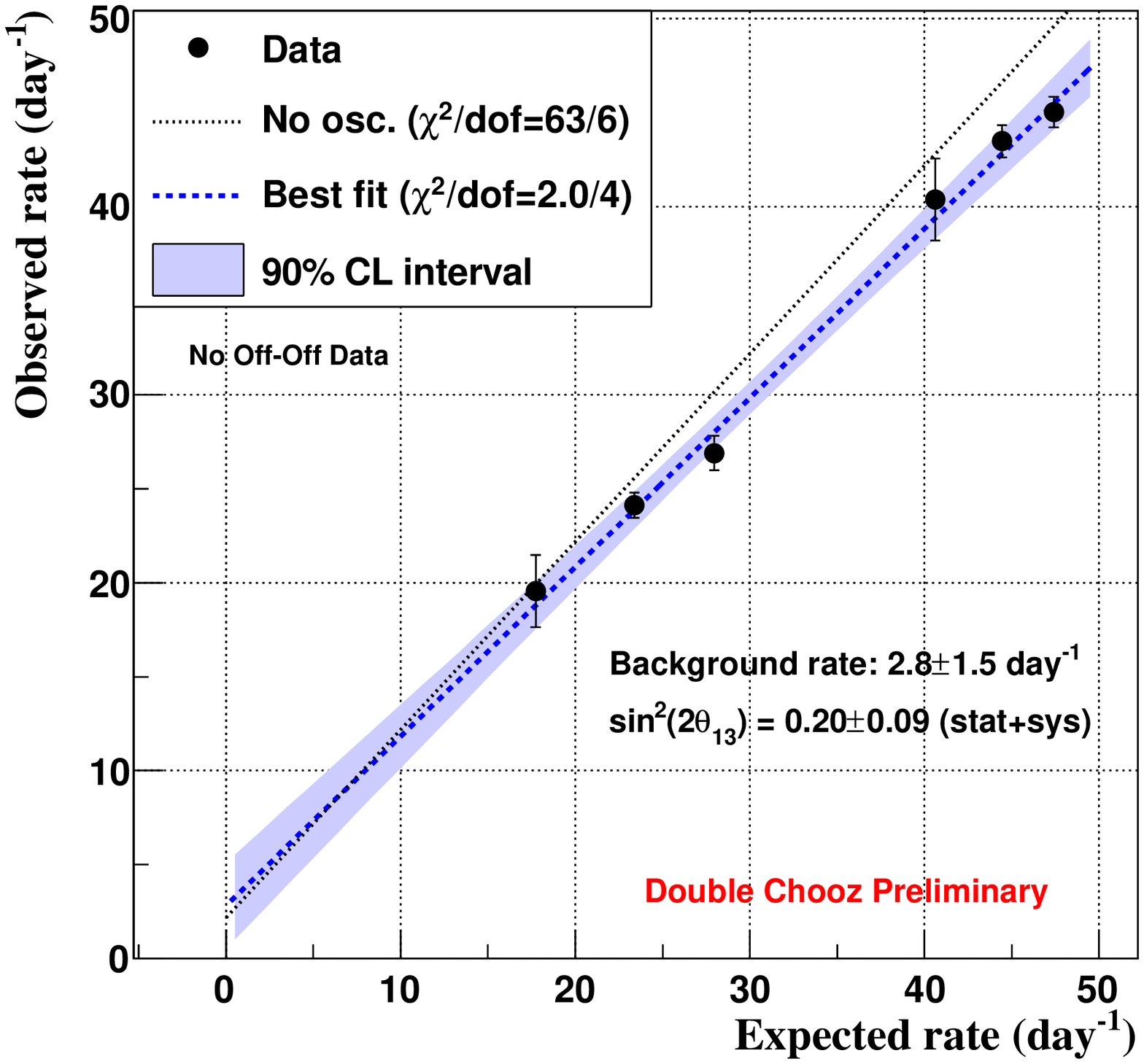}
\includegraphics[width=60mm]{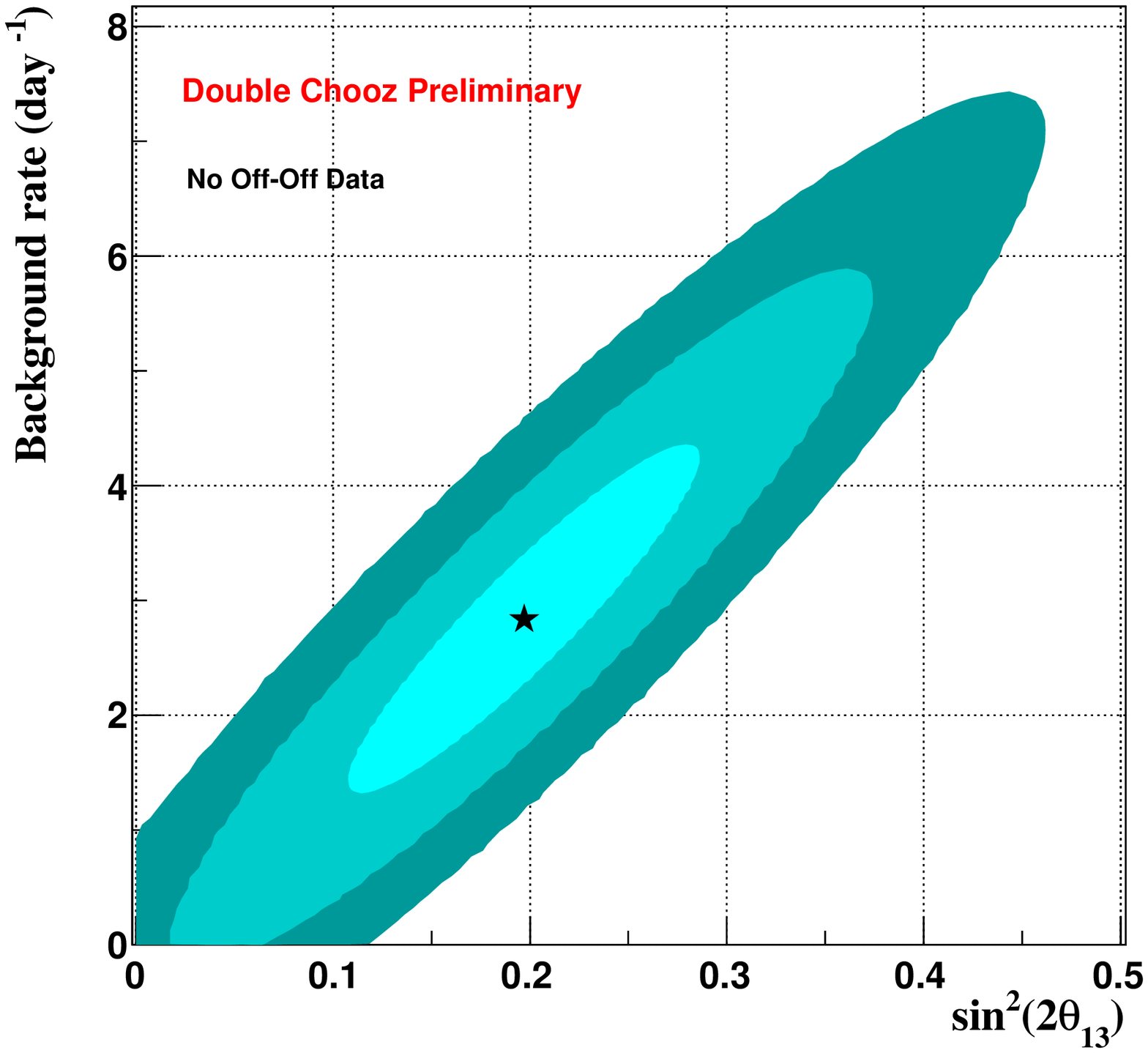}
\end{center}
\caption{\label{fig:fit2} Observed versus expected candidates rate fit without 2-Off reactor data. Left: (sin$^2(2\theta_{13})$,$B$) fit, superimposed to the null oscillation hypothesis. Right: 1, 2 and 3$\sigma$ ($\theta_{13}$,$B$) contour plot.}
\end{figure}

In order to analyze the impact of the 2-Off data sample in this analysis, one can perform the rate fit excluding these data. The results are shown in Fig.~\ref{fig:fit2}: sin$^2(2\theta_{13})$=0.20$\pm$0.09 and $B$=2.8$\pm$1.5 events/day. Although the relative errors on the fit parameters remain almost the same, the central values are deeply affected with respect to the results incorporating the 2-Off data. Thereby, it is concluded that the 2-Off data provides a powerful handle to constrain the central values of $\theta_{13}$ and $B$, even if the limited statistics of this sample does not provide a significant improvement in their precision. It is also worth noticing that the result with no 2-Off data is fully consistent with the one obtained by the Rate-Only analysis in \cite{dc2012}, where the 2-Off sample is not taken into account either. 

\section{Summary and conclusions}

While the oscillation results published by the collaboration in 2011 \cite{dc2011}  and 2012\cite{dc2012} rely on background models derived from reactor-on data, the analysis described in this note is an independent study based on the handle provided by 7.53 days of reactor-off data. A global fit to both $\theta_{13}$ and the total background is performed by analyzing the observed neutrino rate as a function of the non-oscillated expected rate for different reactor power conditions. The outcome of this fit is fully consistent with the one already published by Double Chooz: sin$^2(2\theta_{13})$=0.10$\pm$0.04 and $B$=1.1$\pm$0.5 events/day. As both the published DC results and the current analysis yield almost the same precision, this work stands as a prove of the reliability of the background estimates and the oscillation analysis of the experiment.

\end{document}